\documentclass[a4,titlepage]{article}
\usepackage{epsfig}
\usepackage{amssymb}
\def\beq{\begin{eqnarray}}
\def\eeq{\end{eqnarray}}
\def\beqs{\begin{eqnarray*}}
\def\eeqs{\end{eqnarray*}}

\newcommand{{\SD}}{\rm SD}

\newcommand{\be}{\begin{equation}}
\newcommand{\ee}{\end{equation}}

\def\centeron#1#2{{\setbox0=\hbox{#1}\setbox1=\hbox{#2}\ifdim
\wd1>\wd0\kern.5\wd1\kern-.5\wd0\fi
\copy0\kern-.5\wd0\kern-.5\wd1\copy1\ifdim\wd0>\wd1
\kern.5\wd0\kern-.5\wd1\fi}}
\def\ltap{\;\centeron{\raise.35ex\hbox{$<$}}{\lower.65ex\hbox{$\sim$}}\;}
\def\gtap{\;\centeron{\raise.35ex\hbox{$>$}}{\lower.65ex\hbox{$\sim$}}\;}

\def\lsim{\mathrel{\ltap}}

\begin{document}

\begin{titlepage}
\begin{flushright}
LAPP-EXP-2006-05\\
\end{flushright}

\begin{center}

\vskip 1.2cm

{\LARGE\bf Complementary strategy of New Physics searches in
B-sector.} \vskip 1.4cm

{\large  A.I.Golutvin$^{1,3}$, V.I.Shevchenko$^{2,3}$}\\
\vspace{0.3cm} {\it $^1$CERN, Geneva 23
CH-1211 Switzerland  } \\
{\it $^2$LAPP, 9 Chemin de Bellevue, BP 110 \\
F-74941 Annecy-le-Vieux, France }
 \\
 {\it $^3$Institute of Theoretical and Experimental
Physics
\\B.Cheremushkinskaya 25, 117218 Moscow, Russia }  \vskip 1cm

\end{center}

\begin{center}
{\bf Abstract}

\end{center}

We discuss a possible strategy for studies of a particular scenario
of New Physics (NP) at LHC. The NP is taken to be a). $U$-spin
symmetric, i.e. it does not distinguish $d$ and $s$ quarks; b). it
makes no contribution to the tree processes, but contributes
differently to penguin and box diagrams and c). it does not spoil
the unitarity of CKM matrix. Our analysis is based on comparison of
particular CKM matrix elements, which can be obtained from the
processes dominated by diagrams of different topology. We argue that
the standard formalism of the overall unitarity triangle fit is not
suitable for studies of NP of this kind. We also stress the utmost
importance of lattice computations of some particular set of
hadronic inputs relevant for NP searches.

\end{titlepage}

\section{Introduction}

The main interest of the present day flavor physics is focused
towards searching for possible signals of New Physics (NP) - the
effects which are not taken into account by the Standard Model (SM).
These still hypothetical effects can be roughly divided into two
groups. The first, "quantitative" one consists of effects which are
present in the SM but whose concrete SM prediction deviates from
actual experimental results. A well known example is given by rare
decays strongly suppressed in the SM but expected to be enhanced in
some NP scenarios. Another, "qualitative" group, is formed by the
effects which are not present in the SM at all, like possible
observation of nonconservation of any charge (baryon, electric etc)
strictly conserved in the SM. At the moment there are no clear
indications on possible NP effects of either kind. The strong hope
however is that the situation will change in the nearest future with
the run of LHC. Of prime importance in flavor physics is an analysis
of the CKM mixing matrix. The commonly accepted
parametrization-independent
 language used to discuss the rich physics encoded in CKM matrix is
formalism of the unitarity triangle (UT). For introduction into the
subject and all details the reader is refered to the materials
presented on \cite{ckmfitter,utfit,heavyflavor,ucsd} and to
excellent recent reviews \cite{b,f}.

The issue of NP search in flavor physics context is certainly much
broader than the mere check of CKM matrix unitarity, however precise
it can be. Of course, any inconsistencies in the UT construction
will undoubtedly indicate the presence of physics beyond SM. The
opposite is far from being true - there are many reasonable NP
scenarios which are well compatible with perfect unitarity of CKM
matrix.

There are different possible strategies to study the CKM matrix. The
most popular one, adopted in particular by UTfit and CKMfitter
groups \cite{ckmfitter,utfit} is to use all available experimental
data to overconstrain the triangle. Besides general importance of
this activity the hope is that the procedure will exhibit some
inconsistencies signaling NP effects. Up to now there is an overall
agreement of all constraints (see recent talks
\cite{jampiens,bona}).

However, this approach also has some disadvantages. In our view, the
most important one is the fact that the set of constraints in use is
not fitted to this or that particular NP scenario. On the other
hand, the relevance of this or that observable from the point of
view of its possible NP content strongly depends on what kind of NP
we discuss. Let us explain this point taking as a typical example
$\Delta M_s / \Delta M_d$ ratio. For all scenarios where NP couples
identically to $s$ and $d$ quarks ($U$-spin symmetric NP) this ratio
is not sensitive to NP contributions, since in this case
short-distance functions, even if modified with respect to the SM
predictions for each $\Delta M_d$, $\Delta M_s$ exactly cancel in
the ratio. This pattern is typical for, e.g. constrained minimal
flavor violation (CMFV) NP models (see review of MFV models in
\cite{buras5}). As a result this quantity informs us about ratio of
couplings of $t$-quark to $d$ and $s$-quarks and also about
long-distance $SU(3)$ breaking effects in QCD (see expression
(\ref{lkj}) below), but brings no information about correctness of
the short-distance SM calculation of $\Delta M_{d}$ or $\Delta
M_{s}$ separately. And it is precisely the latter short-distance
piece we are interested in most of all if we are looking for
deviations from the SM at small distances. On the other hand, there
are NP scenarios such as, e.g. MSSM at large $\tan \beta$ (see
\cite{buras4} and references therein) and next-to-minimal flavor
violation \cite{aga} where this is not the case and the ratio under
study is sensitive to NP. Moreover, it is very natural to expect
(and this is our general attitude in the present paper) that NP
contributes differently to the processes of different topology (i.e.
tree and penguin, penguin and box etc). Obviously, this effect can
be lost in comparison of observables of the same topological type.
In view of that an alternative way has been proposed (see, e.g.
\cite{nir1,nir2} and also
\cite{ckmfitter,utfit,buras4,buras2,buras3} and references therein).
Generally speaking, it corresponds to construction of a few a priori
not coinciding unitarity triangles, each extracted from branching
ratios and asymmetries for processes of some particular kind. In
this case any mismatch between these UT's, e.g the so called
"reference UT" \cite{nir2} and "universal UT" (see recent discussion
in \cite{buras4}) would be a clear signal of NP, and, moreover, one
could in principle identify the place (EW penguin sector is among
the most promising ones) where it has come from.

Adopting the basic idea of the latter strategy we address the
following problem. Let us assume the still hypothetical NP is, in
the spirit of next-to-minimal flavor violation scenario: a).
$U$-spin symmetric; b). does not contribute to the tree processes
and c). does not spoil the unitarity of CKM matrix (i.e. we work in
the spirit of next-to-minimal flavor violation scenario). How can we
see NP from global UT fits and what observables are the most
sensitive to NP effects in this particular case?

To answer this question, we analyze theoretical and experimental
(having in mind mostly the LHCb experiment) perspectives for studies
of some CKM matrix parameters which can be extracted from the
processes of different topology and can be sensitive to NP of the
discussed type. It is worth noticing that the mismatch between $\sin
2\beta$ values from $B\to J/\psi K_S$ and from $B\to \phi K_S$ modes
widely discussed in recent literature (see, e.g. \cite{b,fl7} and
references therein) represents exactly a kind of effects we are
interested in. We will also stress the urgent need for new refined
lattice data on hadronic input parameters in order to determine the
product $|V_{ts}V_{tb}^*|$.

The paper is organized as follows. The section 2 is devoted to brief
overview of the existing strategies for CKM matrix analysis, while
our procedure and results are presented in the section 3 and
conclusion in the section 4.

\section{Overview of the standard strategy}

In general one can choose different sets of independent parameters
which enter the basic unitarity relation\footnote{As well as five
other unitarity triangles} \be V_{ud} V_{ub}^{*} + V_{cd} V_{cb}^{*}
+ V_{td} V_{tb}^{*} = 0 \label{tre} \ee It is worth noticing that
the term "independent" is usually used in the literature in mere
algebraic sense, i.e. one assumes no relations between CKM matrix
elements other than those following from the unitarity constraints.
This assumption may be wrong if some more fundamental underlying
structure behind CKM matrix does exist. A common choice for one of
the parameters is \be s_{12} = \lambda = |V_{us}| =
\left\{ \begin{array}{ll} (0.2265 \pm 0.0020) & \mbox{\cite{ckmfitter}} \\
 (0.2258 \pm 0.0014) & \mbox{\cite{utfit}} \\
\end{array}
 \right. \label{vus} \ee
This quantity can be determined with very good accuracy from the
decay mode $K \to \pi l \nu$ with the latter being dominated by tree
level process. The main source of error here is the poor knowledge
of the corresponding formfactor $f_+(0)$, namely, according to
\cite{mes} $\delta |V_{us}|_{f_+(0)} = \pm 0.0018$, $ \delta
|V_{us}|_{exp} = \pm 0.0005$.

 The
interior angles of the triangle (\ref{tre}) are conventionally
labeled as \be \alpha = {\mbox{arg}}\left( -\frac{V_{td}
V^*_{tb}}{V_{ud} V^*_{ub}}\right)\; , \;\; \beta =
{\mbox{arg}}\left( -\frac{V_{cd} V^*_{cb}}{V_{td} V^*_{tb}}\right)\;
, \;\; \gamma = {\mbox{arg}}\left( -\frac{V_{ud} V^*_{ub}}{V_{cd}
V^*_{cb}}\right) \label{e2} \ee The Cabbibo-suppressed angle $\chi$
important for $B_s - {\bar B}_s$ oscillations is defined by \be \chi
= {\mbox{arg}}\left( - \frac{V_{cs}^* V_{cb}}{V_{ts}^*
V_{tb}}\right) \ee and is also of interest. Let us briefly remind
the strategy for $\gamma$. The cleanest way to extract it is from
the interference of the $b\to c{\bar{u}}s$ and $b\to {\bar{c}}us$
transitions (the so called "triangle" approach). Practically, this
corresponds to the study of $B^- \to K^- D^0$ and $B^- \to K^- {\bar
D}^0$ modes with the subsequent analysis of the common final states
for $D$ and ${\bar D}$ mesons decays. One considers $CP$-eigenstates
as final states for $D, {\bar D}$ mesons decays (GLW approach
\cite{gr}) or combines observables from different modes ($B\to
K^*D$, $B\to K D^*$, $B\to KD$, $B\to K^* D^*$) (ADS approach
\cite{atwood}) to overconstrain the system.\footnote{The same
strategy can be applied to the case of $B_c$ mesons, where it has
some theoretical advantages \cite{fl}; however the experimental
statistics becomes the main obstacle there.} Notice that the
interfering diagrams are the tree ones.\footnote{This mode is not
completely NP-safe since in principle the latter can enter through
$D^0 - {\bar D}^0$ mixing.}

 The combined results for $\gamma$ presented in \cite{heavyflavor}
obtained by Dalitz plot analysis \cite{bondar} are given by \be
\gamma = 67^\circ \pm 28^\circ \pm 13^\circ \pm 11^\circ \;
\mbox{[BaBar]} \;\;\; ; \;\; \gamma = 53^\circ \pm 18^\circ \pm
3^\circ \pm 9^\circ \; \mbox{[Belle]} \label{gamma1}\ee where the
errors are statistical, systematic and the error resulting from the
choice of $D$ - decay model. For the discussion of the situation
with $\gamma$ determination in LHCb the reader is referred to
\cite{ruth}.

Using various methods the overall uncertainty in $\gamma$ at LHCb is
expected to be as small as $5^\circ$ in $2 \mbox{fb}^{-1}$  of
running and will eventually reach the level of $1^\circ$ with
increase of statistics.

With the standard assignment for the elements of CKM matrix (see,
e.g., \cite{f}) \be s_{12} = \lambda \; ; \; s_{23} = A\lambda^2 \;
; \; s_{13}\exp(-i\delta_{13}) = A \lambda^3 (\rho - i \eta) \ee to
define the apex of the unitarity triangle \be {\bar\rho} =
\rho\left[1-\frac12 \lambda^2\right] \;\; ;\;\; {\bar\eta} =
\eta\left[1-\frac12 \lambda^2\right] \ee one needs to know at least
two independent quantities out of two sides
$$
R_b = \frac{|V_{ud} V^*_{ub}|}{|V_{cd} V^*_{cb}|} =
\sqrt{{\bar\rho}^2 + {\bar\eta}^2} \; \; ; \;\; R_t = \frac{|V_{td}
V^*_{tb}|}{|V_{cd} V^*_{cb}|} =
\sqrt{(1-{\bar\rho})^2 + {\bar\eta}^2} \\
$$
and three angles $\alpha, \beta, \gamma$ where the latter are
defined by (\ref{e2}). In particular, the authors of \cite{buras3}
analyzed all ten possible strategies, distinguished by the mentioned
choice of two independent parameters out of five from the point of
view of their efficiency in the determination of UT. For example,
our geometrical intuition tells us that it is easier to construct
general non-squashed triangle taking as inputs one of its angles and
adjacent side (because the variations in these parameters are
approximately orthogonal) than taking the same angle and the
opposite side (because the variation in these parameters are
approximately parallel). Numerical simulation done in \cite{buras3}
fully supports this intuition, giving the highest
priority\footnote{In other words, to compute $({\bar\rho},
{\bar\eta})$ with a given precision pair $(\gamma, \beta)$ may be
known with lower accuracy than, e.g. pair $(R_b , \beta)$.} to the
strategies based on combined use of either $(\gamma, \beta)$ or
$(\gamma, R_b)$.

This result is particularly encouraging because the quantities $R_b$
and $\gamma$ define the so called reference UT \cite{nir1,nir2}. The
latter is built from the observables that are expected to be
unaffected by NP, since their dominant contributions come from tree
level processes. Then assuming unitarity of CKM matrix one can
compute from (\ref{tre}) reference values for \be R_t =
\sqrt{1+R_b^2 - 2R_b \cos\gamma} \;\; ; \;\; \cot\beta =\frac{1 -
R_b \cos\gamma}{R_b \sin\gamma} \label{rb}\ee and compare them with
the ones obtained by direct measurements in the processes involving
loop graphs. Any difference could be a hint for a NP signal (see
recent quantitative discussion of this issue in \cite{buras4}).

The elements of CKM matrix which enter the definition of $R_b$ (up
to terms ${\cal O}(\lambda^4)$
$$
R_b =  \left[1-\frac12 \lambda^2\right]\frac{1}{\lambda}
\frac{|V_{ub}|}{|V_{cb}|} $$ are known from semileptonic B-decays.
The recent inclusive update is given by \cite{ckmfitter} $ |V_{cb}|
= (41.79 \pm 0.63)\cdot 10^{-3}$.

Experimental determination of $|V_{ub}|_{incl}$ suffers from
uncertainties, introduced by specific cuts one has to apply in order
to get rid of $b\to c$ background. As for $|V_{ub}|_{excl}$ the main
source of error is lattice uncertainty in calculations of $B \to
\pi, \rho$ form-factors. Up to date results are given by
\cite{heavyflavor} as
$$
|V_{ub}|_{incl} = (4.4 \pm 0.3)\cdot 10^{-3}
$$
and
$$
|V_{ub}|_{excl} = (3.8 \pm 0.6)\cdot 10^{-3}
$$

At the moment the perspectives to increase the accuracy in
experimental determination of $R_b$ up to a few percent level are
unclear. As can be seen from Fig.1 and Fig.2, the errors in $\gamma$
and $R_b$ play a very different role in fixing the angle $\beta$
with some given precision, which is a simple consequence of the fact
that the angle $\alpha$ is close to $90^\circ$ and the triangle is
almost rectangular. The present accuracy in $\beta$ extracted from
the "golden mode" $B \to J/\psi K_S $ is better than $\pm 2^\circ$,
the current world average for $\sin 2\beta$ from tree level decays
provided by \cite{heavyflavor} is (see recent talk \cite{hazumi} and
references therein): \be \sin 2\beta = (0.674 \pm 0.026)
\label{beta} \ee The corresponding penguin contribution to $\beta$
is Cabibbo-suppressed (see, e.g. \cite{fl7}). As shown in Fig.1 an
uncertainty window of $\sim 3^\circ$ for $\beta$ corresponds to an
uncertainty window of $\sim (24\pm 5)^\circ $ for $\gamma$ and
therefore the precise data (\ref{beta}) does not constrain $\gamma$
via (\ref{rb}) strongly enough to make the comparison discussed
above meaningful. On the other hand, since both $\beta$ from
(\ref{beta}) and $\gamma$ from (\ref{gamma1}) are determined from
the processes dominated by tree level decays, we do not expect to
see violation of the second expression from (\ref{rb}) with these
values of the angles. Anyway, the experimental uncertainty in
$\gamma$ and hadronic uncertainties in $R_b$ make (\ref{rb}) not
valuable.

Let us briefly discuss the side $R_t$. The are two ways of
extracting $R_t$ by means of relations not affected by NP
contributions in some scenarios, notably CMFV. These are the
computation of $R_t$ from the first expression in (\ref{rb}) and the
computation from the ratio $\Delta M_d / \Delta M_s$ where, again,
short distance contributions to the box diagrams are
canceled.\footnote{As has been already mentioned, in MSSM at large
$\tan \beta$ the quantity $R_t$ is sensitive to the different Higgs
couplings to $d$ and $s$ quarks.} Concerning the former algorithm,
because of the same geometrical reasons (angle $\alpha$ close to
$90^\circ$) $R_t$ is sensitive to the uncertainty in the angle
$\gamma$ only (see Fig.3). Thus, precise knowledge of $\gamma$ will
constrain $R_t$ effectively.
In the latter approach one  obtains the ratio \be
\frac{|V_{td}|}{|V_{ts}|} = \xi \sqrt{ \frac{m_{B_s}}{m_{B_d}}
\frac{\Delta M_d}{\Delta M_s} } \label{lkj} \ee with the
nonperturbative parameter\footnote{$\xi =1$ in case of exact flavor
$SU(3)$.} \be \xi^2 = \frac{{\hat B}_{{B_{s}}} f^2_{{B_{s}}}}{{\hat
B}_{{B_{d}}} f^2_{{B_{d}}}} \label{pol} \ee The typical error of
current lattice simulations of $\xi$ is estimated as $6\%$ (see
\cite{okamoto} and recent analysis in \cite{bfl}). Since up to
${\cal O}(\lambda^4)$
$$
R_t = \frac{\xi}{\lambda} \sqrt{\frac{\Delta M_d}{\Delta
M_s}}\sqrt{\frac{m_{B_s}}{m_{B_d}}}\left[ 1 - \lambda \xi \cos\beta
\sqrt{\frac{\Delta M_d}{\Delta M_s}}\sqrt{\frac{m_{B_s}}{m_{B_d}}} +
\frac{\lambda^2}{2} \right]
$$
then having at our disposal recent CDF results \cite{cdf} (see
(\ref{opiu})) we can straightforwardly extract for the mean value $
R_t = 0.92 $ with the uncertainty dominated by $\xi$.

Let us summarize this part. Suppose we would be able to measure
$R_b$ and $\gamma$ with some very high precision. This defines the
position of the UT apex which is universal as soon as NP does not
contribute to tree processes $R_b$ and $\gamma$ have been extracted
from. Let us also assume that we get $R_t$ from $\Delta M_d / \Delta
M_s$ and $\beta$ from $B\to J/\psi K_S$, and these observables
perfectly agree with $R_b$ and $\gamma$ via (\ref{rb}) (i.e. the UT
apex defined from $R_t$ and $\beta$ coincides with the one
found\footnote{Of course, any other pair can actually be used, see
discussion above.} from $R_b$ and $\gamma$). Does this fact mean
dramatic shrinking of NP parameter space? Not at all: for NP
scenarios with $U$-spin invariance and without sizeable NP mixing
effects\footnote{NP physics contributions in the box diagrams could
in principle affect both $\gamma$ and $\beta$ via $D_0 - {\bar D}_0$
and $B_0 - {\bar B}_0$ mixings.} this coincidence is trivial and
brings no any information about the parameter space. One can tell
that UT is simply too rough tool to see NP of this kind. In other
words, the precise knowledge of $\xi$ is important in this case to
calibrate the lattice, but not to find the NP.

\section{Direct comparison of CKM matrix elements from different processes}

In what follows we are going to explore a complementary strategy
whose essence is the comparison of values of CKM matrix elements,
obtained from processes with dominant contributions coming from
diagrams of essentially different topology. Again one can consider
angles and sides in this respect. We are interested in observables,
corresponding to the processes whose dominant contributions come
from topologically different diagrams, namely:
\begin{itemize}
\item radiative penguin in decay modes $B\to K^* \gamma$, $B_s \to
\phi \gamma$ and $B\to (\rho, \omega) \gamma$, $B_s \to {\bar K}^*
\gamma$, for $s$ and $d$ quarks, respectively
\item
oscillations of neutral $B^0$ and $B_s$ mesons with the dominant
contribution given by box diagram, resulting in the mass shifts
$\Delta M_s$, $\Delta M_d$
\item tree and strong penguin interference in $B$ decays
into 2-body final states made of light hadrons $\pi$, $K$, $\rho$
and mixing relevant for the angle $\alpha$ determination
\end{itemize}
For the first and the second mode the object of our interest is the
product $|V_{tb}^*V_{ts}|$. For the third mode we confine our
attention to the angles $\alpha$, $\beta$ and $\chi$.

The reference values for these quantities are defined from tree
level processes, since we adopt the usual assumption that they are
free from NP pollution. One can make use of the "tree level
definition" for $|V_{tb}^*V_{ts}| \leftrightarrow
|V_{tb}^*V_{ts}|_{tree}$, which up to terms ${\cal O}(\lambda^4)$
reads \be |V_{ts}V_{tb}^*|_{tree} = |V_{cb}| \left[ 1 -
\frac{\lambda^2}{2} (1- 2 R_b \cos\gamma)\right] \label{trtr} \ee We
have already discussed the corresponding numerical values and their
uncertainties. Plugging them in, we get \be |V_{ts}V_{tb}^*|_{tree}
= (41.3\pm 0.8)\cdot 10^{-3} \ee As for the angle $\alpha$, its
reference value is given by \be \alpha_{tree} = \pi - \beta - \gamma
\label{alphatree} \ee
 where the extraction of $\beta$ and $\gamma$ from tree level processes is described
above. As for the angle $\chi$, there are no experimental
constraints on it at the moment. The SM prediction is $|\chi|
\approx 0.02 \div 0.04$.

\subsection{Analysis of  $|V_{tb}^*V_{ts}|$.}

Let us start with the analysis of  $|V_{tb}^*V_{ts}|$. Values of
these elements of the CKM matrix must exactly coincide in the SM,
regardless of the way they are extracted. On the other hand, lack of
such coincidence will be a definite signal of NP, contributing
differently to these different types of processes. Qualitatively,
one can consider ratios of the following kind \be \zeta_{q,V}^{(1)}
= \frac{|V_{tb}^*V_{tq}|_{\Delta M_q
\hphantom{111}}}{|V_{tb}^*V_{tq}|_{B \to V \gamma}} \; ; \;
{\zeta}_{q,V}^{(2)} = \frac{|V_{tb}^*V_{tq}|_{B \to V
\gamma}}{|V_{tb}^*V_{tq}|_{tree}} \; ; \; {\zeta}_{q,V}^{(3)} =
\frac{|V_{tb}^*V_{tq}|_{tree}}{|V_{tb}^*V_{tq}|_{\Delta M_q }}
\label{xi2} \ee
 where $q=d,s$ and $V$ stands for $K^* , \phi, \rho, \omega$.
 Thus we have three ways to extract the product $|V_{tb}^*V_{tq}|$ of CKM matrix elements:
via expression (\ref{gamma}) from the process dominated by the
radiative penguin diagram,  via expression (\ref{deltam}) from the
process dominated by the box diagram, and via  (\ref{trtr}) from the
reference tree level processes. It is obvious that by construction
one has \be \zeta_{q,V}^{(1)}\cdot \zeta_{q,V}^{(2)} \cdot
\zeta_{q,V}^{(3)} \equiv 1 \label{basic} \ee In the SM however much
more restricted condition has to be fulfilled: \be \zeta_{q,V}^{(1)}
= \zeta_{q,V}^{(2)} = \zeta_{q,V}^{(3)} =1 \ee It is convenient to
present a set of three numbers $\{\zeta_{q,V}^{(1)},
\zeta_{q,V}^{(2)}, \zeta_{q,V}^{(3)}\}$ as a single point on the
ternary coordinate system with $\log \zeta_{q,V}^{(i)}$ as an
(algebraic) distance from the $i$-th axis. Then the SM case
corresponds to the only point on this diagram - its origin, while
any deviation from it is a hint to NP.

The analysis of ratios of the mass shifts ${\Delta M_d} / {\Delta
M_s}$  and branchings $Br(B\to\rho\gamma)/Br(B\to K^* \gamma)$
widely discussed in the recent literature \cite{bfl,bz,ali1} deals
in our language not directly with the quantities
$\zeta_{q,V}^{(i)}$, but with their ratios like $\zeta_{s,
K^{*}}^{(1)} / \zeta_{d, \rho}^{(1)}$. The important advantage of
these ratios is the improved accuracy of their theoretical
determination, especially from the point of view of hadronic
uncertainties. However the price to pay is high - the short distance
factors which could contain contributions of NP are canceled in
these ratios. In logarithmic coordinates it corresponds to a
parallel translation, which could miss a considerable piece of NP,
which is clearly seen from the analysis of \cite{bfl}. In short,
$\zeta_{s, K^{*}}^{(1)} = \zeta_{d, \rho}^{(1)} = 1$ implies
$\zeta_{s, K^{*}}^{(1)} / \zeta_{d, \rho}^{(1)} = 1$, but not vice
versa.

Generally speaking, it is meaningless to look for (short-distance)
deviations from the SM predictions if one  has no quantitative
knowledge what the latter actually are. Therefore as soon as we are
discussing absolute values of mass shifts, widths etc, these short
distance parameters should be determined and not just canceled in
the ratios. Corresponding loss in an accuracy for hadronic
contributions is perhaps inevitable. Anyway we are stressing that
one has to deal with this "less accurate" low energy hadronic inputs
if one tries to capture the short-distance effects of NP. For
example, it is meaningless, in our view, to consider soft quantities
as free parameters to fit observable branching ratios. Any possible
NP induced difference between, e.g. the SM prediction for $Br (B\to
V \gamma)$ and actual experimental result would be just hidden
inside such "extracted from experiment" $|\xi_\bot^{(K^*)}(0)|$,
which is clearly unacceptable. Simply speaking, to discuss
deviations from the SM prediction we have first to know the
latter.\footnote{As is correctly pointed out in \cite{utfit} "a
model-independent UT analysis beyond the SM cannot be carried out
without some {\it a priori} theoretical knowledge of the relevant
hadronic parameters."}

In principle, one can discuss five expressions of the kind
(\ref{basic}), corresponding to the following choices for $(q, V)$:
$(s, K^{*})$,  $(s, \phi)$,  $(d, \omega)$, $(d, \rho)$,  $(d,
{\bar{K}}^{*})$. However all these channels have universal
short-distance structure, while the long-distance contributions are
related to each other by $SU(3)$ flavor arguments. The optimal
strategy therefore seems to choose just one particular case, which
we take to be  $(s, K^{*})$ in the rest of the paper. The results
for the other ones could provide important cross-checks (like, e.g.
$|V_{td}|/|V_{ts}|$ ratio), but presumably no new information about
a NP content of (\ref{basic}).

We are using the standard SM expressions for the decay rate for $B
\to K^* \gamma$ and the mass difference $\Delta M_s$. The former can
be written as \cite{ali1,bosch,ff2}: \be \Gamma(B\to K^* \gamma) =
\frac{G_F^2 \alpha m_B^3 m_b^2}{32\pi^4}(1-r)^3 |a_7(\mu)|^2
|\xi_\bot^{(K^*)}(0)|^2 |V_{ts}V_{tb}^*|^2 \label{gamma} \ee In the
above expression $r = m_{K^*}^2 / m_B^2 $, $m_b$ stands for the pole
mass of $b$-quark, $a_7(\mu) = C_7^{(0)} + A^{(1)}(\mu)$ is an
absolute value of the corresponding short-distance function
including Wilson coefficient $C_7^{(0)}$, hard scattering
contributions and annihilation corrections. The detailed computation
of this function at the next-to-leading order can be found in the
cited papers. Notice that we omit terms of the order of
$m_s^2/m_b^2$. The factor $|\xi_\bot^{(K^*)}(0)|$ differs from the
corresponding form-factor $T_1^{B\to K^*}(0)$ by ${\cal
O}(\alpha_s)$ corrections; according to \cite{beneke} numerically
one has $|\xi_\bot^{(K^*)}(0)| \approx 0.93 \cdot T_1^{B\to
K^*}(0)$.

The expression for $\Delta M_s$ reads as follows:
 \be
\Delta M_{s} = \frac{G_F^2}{6\pi^2} \eta_B [M_W^2 F_{tt}] m_{B_{s}}
({\hat B}_{{B_{s}}} f^2_{{B_{s}}}) |V_{ts}V_{tb}^*|^2 \label{deltam}
\ee where $\eta_B$ is calculable short-distance QCD factor, while
the $m_t/M_W$-dependent factor $F_{tt}M_W^2$ has come from
calculation of the box diagram (\cite{vis}, see also \cite{bbl,il}).

According to our strategy we invert the expressions (\ref{gamma})
and (\ref{deltam}) to the following form: \be |V_{ts}V_{tb}^*|_{B
\to K^* \gamma} = \frac{4\pi^2}{|a_7(\mu)|} \cdot \sqrt{\frac{2
\Gamma(B\to K^* \gamma) }{G_F^2 \alpha m_B^5 (1-r)^3}} \cdot \left[
\frac{1}{|\xi_\bot^{(K^*)}(0)|} \left( \frac{m_B}{m_b}\right)
\right]\label{vvg}
 \ee and \be |V_{ts}V_{tb}^*|_{\Delta M_s} =
\frac{\pi}{\sqrt{\eta_B F_{tt}}} \cdot \sqrt{\frac{6\Delta
M_s}{G_F^2 M_W^2 m_{B_s}^3 }} \left[ \frac{m_{B_s}}{f_{B_s}
\sqrt{{\hat B}_{{B_{s}}}}} \right] \label{vvbox}
 \ee

The structure of the above expressions is clear. The first factors
in the r.h.s. are the short-distance SM contribution, which have to
be calculated analytically. These are just numbers of order 1 and it
is assumed that we have reliable theoretical control of this part.
The typical accuracy of these factors is better than 5\%. The second
factors (the square roots) are composed from experimentally
measurable quantities. The error in these factors is dominantly
experimental and is currently at the 5\% level for (\ref{vvg}) and
1-2 \% level for (\ref{vvbox}) . The third factors (in the square
brackets) encode information about soft QCD contributions (and
related problem of $b$ quark pole mass $m_b$) for which we have no
systematic approach of studying. The main hope here is focussed on
the lattice simulations.\footnote{Notice that we have included the
$B$-meson mass $m_B$ to both factors in (\ref{vvg}) to provide
normalization. Indeed, it can be understood as being taken from real
experiment, on the other hand since any reliable lattice simulation
must be correctly normalized to this experimental value of the mass,
the treatment of $m_B$ as a lattice output should make in fact no
difference.} The uncertainty of currently available data can be
conservatively estimated as 10-20 \%. The use of (\ref{vvg}),
(\ref{vvbox}) as probes for NP entirely depends on improvement in
the determination of these hadronic factors.

The quantities of our interest are $T_1^{B \to K^* }(0)$ and
$f_{B_{s}} \sqrt{{\hat B}_{B_{s}}}$.The reader is referred to the
papers \cite{abs} - \cite{bec49}  and the papers \cite{bz1,bec50}
for lattice and sum rule determination of $T_1^{B\to K^* }(0)$,
respectively. The corresponding values are in the range $0.2 - 0.4$.
The relevant references for $f_{B_{s}} \sqrt{{\hat B}_{B_{s}}}$ are
given by papers \cite{lel} - \cite{schier}. Looking at the data one
can see that there is no clear agreement between, e.g. lattice
computations and light-cone sum rules results. Moreover, the errors
given by the authors of the cited lattice papers are mostly
statistical ones. The procedure of correct treatment of systematic
errors in this case is not yet known. In fact, the same is true for
the sum rule calculations. In general, precise determination of
$T_1^{B\to K^* }(0)$ on the lattice is very difficult and
reliability of the calculations done so far is debatable (see
\cite{bec} and recent discussion in \cite{wing}). However the utmost
importance of this measurement, which hopefully will be done in the
nearest future on new improved lattices in unquenched case cannot be
overestimated.

Thus, having no better strategy at the moment, we will be
conservative in our error treatment and take for the input value of
$f_{B_{s}} \sqrt{{\hat B}_{B_{s}}}$
$$  f_{B_{s}} \sqrt{{\hat
B}_{B_{s}}} = (280 \pm 40) \;\mbox{MeV}
$$
while we also consider three sets of possible values for $T_1^{B\to
K^* }(0)$, where the errors correspond to those reported in the
cited papers:
\begin{eqnarray*} {\mbox{Set A}} \;\; : \;\;\; T_{1A}^{B\to K^*
}(0) = (0.25 \pm 0.05) \\
{\mbox{Set B}} \;\; : \;\;\; T_{1B}^{B\to K^* }(0) = (0.30 \pm 0.05)\\
{\mbox{Set C}} \;\; : \;\;\; T_{1C}^{B\to K^* }(0) = (0.35 \pm 0.05)
\end{eqnarray*}
The ultimate goal should be to reach the accuracy of the lattice
computations comparable to the accuracy of the r.h.s. of
(\ref{vuss}).

Finally, let us recall the experimental data for $B_s$ meson levels
splitting $\Delta M_s $ \cite{cdf} and branching ratios for the
decay $B\to K^* \gamma$. They are given by \be \Delta M_{s} =
[17.33^{+0.42}_{- 0.21} (stat) \pm 0.07 (sys)] {\mbox{ps}}^{-1}
\label{opiu} \ee and
\begin{eqnarray}
Br(B^-\to K^{*-} \gamma) = (4.25 \pm 0.31 \pm 0.24)\cdot 10^{-5}
\;\;\;\;\; \mbox{\cite{bbb1}} \\
Br(B^-\to K^{*-} \gamma) = (3.87 \pm 0.28 \pm 0.26)\cdot 10^{-5}
\;\;\;\;\;  \mbox{\cite{bbb2}}
\end{eqnarray}
For the life time of $B^-$ meson we use the value $\tau = (1.652 \pm
0.014) \;\mbox{ps}$  \cite{pdg}, while the masses (in MeV) are given
by \cite{pdg}
$$
m_B = (5279.0 \pm 0.5) \; ; \;\; m_{B_s} = (5367.5 \pm 1.8) \; ;
\;\; m_{K^*} = (891.66 \pm 0.26)
$$
Other short-distance inputs are collected in the Table 1.

We have all input data now to estimate the ratios $\zeta_{s, K^*}$.
According to the three choices of numerical value for the
form-factor $T_1^{B\to K^*}(0)$ we get three sets of $\zeta_{s,
K^*}^{(i)}$. The results are presented in Table 2. For graphical
presentation one can use planar ternary coordinates where the
constraint $\sum_{i=1}^3 \log \zeta_{s,K^*}^{(i)} = 0$ is satisfied
automatically. Each solution is represented by a single point on
this plane with the distance from the $i$-th axis to the point given
by $\log \zeta_{s,K^*}^{(i)}$. It is taken positive for two axes
forming an angle the point belongs to and negative for the remaining
distant axis. With this rule each point on the plane satisfies the
constraint (\ref{basic}). Some sample result for the case [44]-B is
shown on Figure 4. Notice, that the bars correspond to $1 \sigma$
deviation in $\zeta_{s,K^*}^{(i)}$, not in $\log
\zeta_{s,K^*}^{(i)}$. The fact that they cross the corresponding
axes means less than $1\sigma$ deviation of the actual result from
the SM prediction. The origin of this ternary coordinate system
corresponds to $\log \zeta_{s,K^*}^{(i)} = 0$ for all $i$, which is
the SM solution.

The main qualitative conclusion is perhaps not surprising: with the
reasonable choice of parameters we observe no evidence for NP within
error bars. There are two optimistic remarks however. The first is
that our errors are very conservative and significant reduction of
at least some of them is foreseen in the nearest future. Secondly,
the errors in the Table 2 are not independent. There are two sorts
of correlations. The first is the uninteresting "kinematical" one,
following from the constraint (\ref{basic}). The second pattern
corresponds to the error correlation for lattice simulations of
$T_1^{B\to K^*}(0)$ and $f_{B_{s}} \sqrt{{\hat B}_{B_{s}}}$. So far
these two inputs have been measured independently, by different
lattice groups and within different procedures. Correspondingly, the
errors shown in the Table 2 are also treated as independent. On the
other hand, it is reasonable to expect an error reduction for the
simultaneous calculation of $T_1^{B\to K^*}(0)$ and $f_{B_{s}}
\sqrt{{\hat B}_{B_{s}}}$ and we call the attention to importance of
such simulation, using the same framework (lattice action, chiral
extrapolation procedure etc) and uniform error treatment. It is
reasonable to expect that this would result in a better accuracy,
first of all for the quantity $\zeta_{s, K^*}^{(1)}$. Speaking
differently, if one assumes no NP (i.e. $\zeta_{s, K^*}^{(1)} = 1$)
one is to get \be \frac{m_b |T_{1}^{B\to K^* }(0)|}{f_{B_{s}}
\sqrt{{\hat B}_{B_{s}}}} = 778 \cdot \left[ Br (B\to K^* \gamma)
\right]^{1/2} \label{vuss} \ee where the uncertainty in the
numerical factor $778$ is of order $~5\%$ and is mostly
theoretical.\footnote{The uncertainty in experimental value of
$\Delta M_s$ is small.} This SM prediction demonstrates the level of
precision the lattice computations must reach in order to make
reliable conclusion about NP based on the lattice results. We
consider the check of (\ref{vuss}) on the lattice as a task of
primary importance.

\subsection{Analysis of the angle $\alpha$}

The angle $\alpha$ can be extracted from the two-body decay modes of
$B$ into light hadrons $\pi, \rho$ and $K$ (see recent review
\cite{m}).
 From the theoretical point of view the
best channel seems at present to be  $B\to \rho\rho$
\cite{gronau,gr1}. The most promising channel for $\alpha$ at LHCb
however is $B\to \rho\pi \to \pi\pi\pi$ \cite{gro5,des}. The basic
idea of the analysis \cite{sq} is to study the interference of the
tree amplitude proportional to the weak phase factor $e^{i\gamma}$
from $V_{ub}^*V_{ud}$ and the penguin amplitude proportional to
factor $e^{-i\beta}$ from $V_{tb}^*V_{td}$. Writing down also the
amplitudes for $CP$-conjugated modes and imposing isospin relations,
one can fit four amplitudes, four strong phases and one weak phase
from 11 observables (see details in \cite{des}). The expected
uncertainty in $\alpha$ of LHCb is about $10^\circ$ in one year of
running \cite{ruth}. It can be mentioned that the recent result
presented by BaBar collaboration \cite{bbbb} for $\alpha$ from $B\to
\rho\pi$ channel is \be \alpha = (114 \pm 39)^{\circ} \ee while the
data uncertainty for $B\to \rho\rho$ mode is $\pm 13^\circ$
\cite{bbb1}. The above analysis assumes no electroweak penguin
contributions. According to the estimates \cite{gr1}, $\delta
\alpha_{EWP} = -1.5^\circ$. The isospin breaking effects controlled
by parameter $(m_d - m_u)/\Lambda_{QCD}$ are expected to be of the
same order of magnitude.

For $\alpha$ defined as an argument of the amplitude ratio one gets
(see details in, e.g. \cite {f}) \be 2\alpha_{eff} = {\mbox{arg}}
\left[-e^{-i\theta_{12}} \frac{A({\bar B}\to f)}{ A(B \to f)
}\right] = {\mbox{arg}} \left[- e^{-i\theta_{12}} \frac{e^{-i\gamma}
- r e^{i\theta + i \delta\alpha}}{ e^{i\gamma} - r e^{i\theta - i
\delta\alpha}}\right]\ee where $\theta_{12}$ is the $B_0 - {\bar
B}_0$ mixing angle, $\theta$ is strong penguin phase, $r$ is an
absolute value of penguin-to-tree ratio and $\delta \alpha$ is
possible weak NP penguin phase.\footnote{The above assignment is
self-consistent for $r \lsim 1$, where terms in $r$ which are
nonlinear in penguin amplitudes can be neglected.} In the absence of
penguins, i.e. if $r=0$ and if $\theta_{12} = 2\beta$ (as in the
SM), one gets $\alpha_{eff} = \alpha_{tree}$ with $\alpha_{tree}$
defined by (\ref{alphatree}). It is worth stressing (see early
discussion of related issue in \cite{nir1}) that for the discussed
scenario the corresponding NP phase shift $\delta \alpha$  is to
coincide up to a sign with that to the angles $\beta$ and $\chi$:
\be \delta \beta _{NP} = \beta_{(B\to \phi K_S)} - \beta_{(B\to J /
\psi K_S)} =  - \delta \alpha = \delta \chi \label{uh} \ee due to
the assumed $U$-spin invariance.\footnote{See discussion of the
related issue in supersymmetric context in \cite{barb}.} Also it has
to be noticed that the box diagram corresponding to the $B^0 - {\bar
B}^0$ mixing contributes identically to the discussed decay modes
and its contribution to the phase (with a possible NP part) is
canceled in (\ref{uh}). Certainly beyond the SM one could have
$\theta_{12} \neq 2\beta$, but this phase shift may have no direct
relation to the discussed shift $\delta \alpha = - \delta \beta$,
resulted from the penguin process. Thus we are left with the only NP
contribution from the penguin-mediated decay (with respect to the
tree level one). We see that the ability to extract $\delta \alpha$
from experiment (i.e. from $\alpha_{eff}$) crucially depends on the
value of $r$, since given experimental uncertainty in $\alpha_{eff}$
corresponds to larger uncertainty in $\delta \alpha$ smaller the
ratio $r$ is. The combined fit of the data for $B\to \rho \pi$ and
other modes (notably $B_0 \to K^*_0 \rho_0$) taking into account
nonzero penguin NP phase $\delta \alpha$ is being performed and will
be reported elsewhere. Here we would like to notice that the
experimental accuracy of $\delta \beta$ is currently limited by the
statistics of $B\to \phi K_S$ decay and recent update for $\sin
2\beta$ from penguin decay modes as given by \cite{hazumi} is $ \sin
2\beta|_{peng} = 0.58 {+0.12 \atop -0.09} \pm 0.13 $ which
correspond to about $15^\circ$ uncertainty window in the angle
$\beta$. It is worth mentioning that this penguin-dominated mode
would not allow to get $\sin 2\beta$ (and hence $\delta \beta$) with
competitive precision at LHCb, since the latter is expected to be
about 0.2 in $2 {\mbox{fb}}^{-1}$ of running \cite{ruth}. Higher
accuracy should be possible for Super-B factories.

\section{Conclusion}

The standard approach to study CKM matrix is to overconstrain the UT
using all available experimental information. However not all
constrains on the $(\rho, \eta)$ plane are sensitive to NP, at least
if the latter is taken in the form of next-to-minimal flavor
violation. Some (such as $\Delta M_d / \Delta M_s$) do not
distinguish the SM from many NP scenarios just by construction,
while others (such as relation (\ref{rb})) are insensitive to NP
because of the specific profile of the UT ($\alpha$ close to
$90^\circ$). In this sense there are two possible points of view
regarding the fact that up to now all constraints on $(\rho, \eta)$
plane agree with each other. The first one is that there are no
sizeable NP effects seen in flavor physics. The second one is that
UT is simply not suitable for the purpose (since NP is not present
in the angles determined from the tree processes and could also
cancel from the sides) and the room for manifestations of NP in
b-physics observables is in fact not so small (because the
uncertainties are still rather large). Following the latter
attitude, we have discussed in this paper a complementary analysis
of the data on the CKM matrix elements. Its key feature is the use
of CKM matrix elements ratios which are sensitive to NP provided it
contributes differently to the processes of different topology. In
this sense the quantities $\zeta_{s, K^*}^{(i)}$ are different from
the ratios like $\Delta M_d / \Delta M_s$ since the short-distance
part is kept in the former. Moreover, since we have more than one
choice for observables a given CKM matrix element is extracted from,
we could have relations of the form (\ref{basic}), leaving
unconstrained more than one degree of freedom. Thus the lattice
simulations must match {\it several} hadronic inputs simultaneously
(and not just {\it one}). This, we believe, will allow to reduce the
corresponding errors and consequently to make the proposed probes
more sensitive to the NP.\footnote{The importance of correlating of
$\Delta F =1$ and $\Delta F = 2$ processes is stressed in another
respect in \cite{aga}.} Speaking differently, one of our main
messages to the lattice community is that the importance of further
reducing uncertainties in the ratio $\xi$ is limited with respect to
the calculation of hadronic inputs entering the definition of
$\zeta$'s since the latter are more sensitive to NP than the former.

Concerning the determination of UT angles which are free from
lattice uncertainties, we advocate the importance of estimates of
the angle $\delta \alpha$ corresponding to the penguin amplitude
extracted from $B\to \rho \pi$ and other modes (and hence subject of
possible NP shifts). The accuracy of such a comparison can be
comparable or better at LHCb than for $\sin 2\beta$ extracted from
$B \to J/\psi K_S$ and $B\to \phi K_S$ modes, while the physical
meaning is the same; any discrepancy between these values would
undoubtedly indicate NP.

In principle, nothing prevents one to include the discussed
quantities $\zeta^{(i)}_{q,V}$ and $\delta (\alpha, \beta, \chi)$
into the global fit of the CKM matrix. It is clear that one gets
essentially no new information in this way, since we deal with the
same experimental observables the standard fitting procedure does.
We feel, however, that careful analysis of the proposed observables
provides an alternative and transparent way of looking at NP
effects. This strategy can become useful in the nearest future when
LHC data will improve the accuracy of our knowledge of the CKM
matrix elements dramatically.

\bigskip

{\bf Acknowledgements}

The authors are grateful to R.Fleischer, R.Forty, J. van Hunen,
A.Kaidalov, T.Nakada, M.Vysotsky and L.Okun for useful discussions.
The work was supported by the grant INTAS-05-103-7571 and partly by
the grant RFFI-06-02-17012a and grants for support of scientific
schools NS-843.2006.2; contract 02.445.11.7424/2006-112. One of the
authors (V.Sh.) thanks the staff of Laboratoire d' Annecy-le-vieux
de Physique des Particules (LAPP) and personally Prof. Bolek
Pietrzyk for his kind hospitality. V.Sh. expresses his gratitude to
the "Dynasty" foundation and ICFPM for financial support. The
support from the President's grant MK-2005.1530.02  is also
acknowledged.


\newpage

\begin{table}
\caption{Short-distance quantities from the definition of $\kappa$.}
\begin{center}
\begin{tabular}{|l|l|l|l|}
\hline Quantity & Mean & Error & Reference \\
\hline
$\eta_B$ & 0.551 & 0.008 & \cite{bbl} \\
\hline
$F_{tt}$ & 2.35 & 0.06 & \cite{il} \\
\hline
${\bar m}_t(m_t),$ GeV & 164.7 & 2.8 & \cite{tevex}\\
\hline
${\bar m}_b(pole),$ GeV & 4.65 & 0.10 & \cite{ali1} \\
\hline $|C_7^{(0)} +
A^{(1)}(\mu)|^2$ & $0.16$ & 0.01 & \cite{ali1} \\
\hline
$m_W,$ GeV & 80.40 & 0.03 & \cite{pdg}\\
\hline
\end{tabular}
\end{center}
\end{table}

\begin{table}
\caption{Numerical results for $\zeta^{(i)}_{s,K^*} / \Delta
\zeta^{(i)}_{s,K^*} $. The abbreviation [43]-A corresponds to the
branching ratio for $B\to K^* \gamma$ from [43] and the Set A choice
for $T_1^{B\to K^*}(0)= 0.25\pm 0.05$, and analogously for other
columns.}
\medskip
\begin{center}
\begin{tabular}{|l|l|l|l||l|l|l|}
\hline  & [43]-A& [43]-B & [43]-C & [44]-A & [44]-B & [44]-C \\
\hline $\zeta^{(1)}_{s,K^*}$ & $0.81 / 0.21$ & $0.97 / 0.23$ & $1.13
/ 0.25$ &$ 0.85 / 0.22$ & $1.02  /  0.24$
&$ 1.18  /  0.26$ \\
\hline $\zeta^{(2)}_{s,K^*}$ & $1.12  /  0.24$ & $0.94  /  0.18$ &
$0.80  /  0.13$ & $1.07  /  0.23 $& $0.89  /  0.17$
&$ 0.77  /  0.12$ \\
\hline $\zeta^{(3)}_{s,K^*}$ & $1.10  /  0.17 $&$ 1.10  /  0.17$&
$1.10  /  0.17$& $1.10  /  0.17$& $1.10
 /  0.17$&$ 1.10  /  0.17$ \\
\hline
\end{tabular}
\end{center}
\end{table}

\newpage

\begin{figure}[!t]
\epsfxsize=10cm \epsfbox{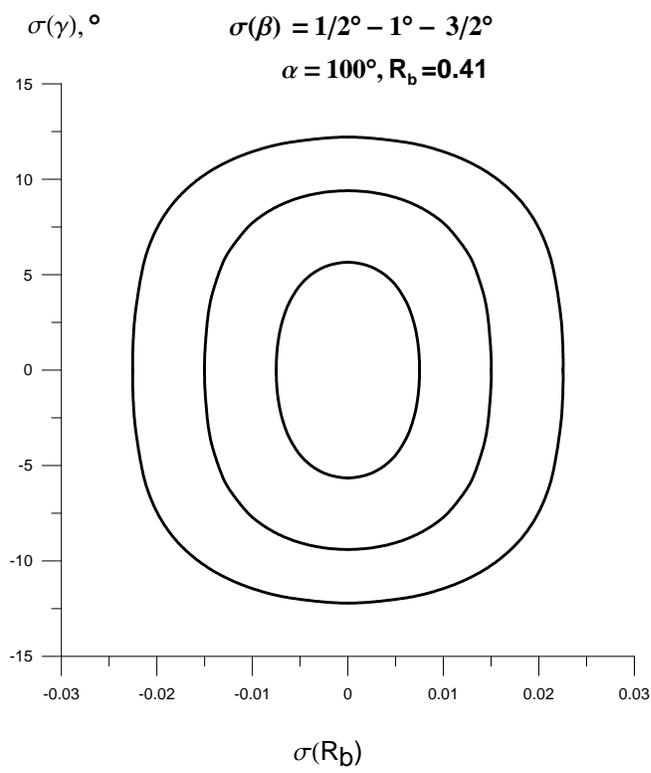} \caption{Error propagation
corresponding to the second expression from (\ref{rb}). }
\end{figure}

\begin{figure}[!t]
\epsfxsize=10cm \epsfbox{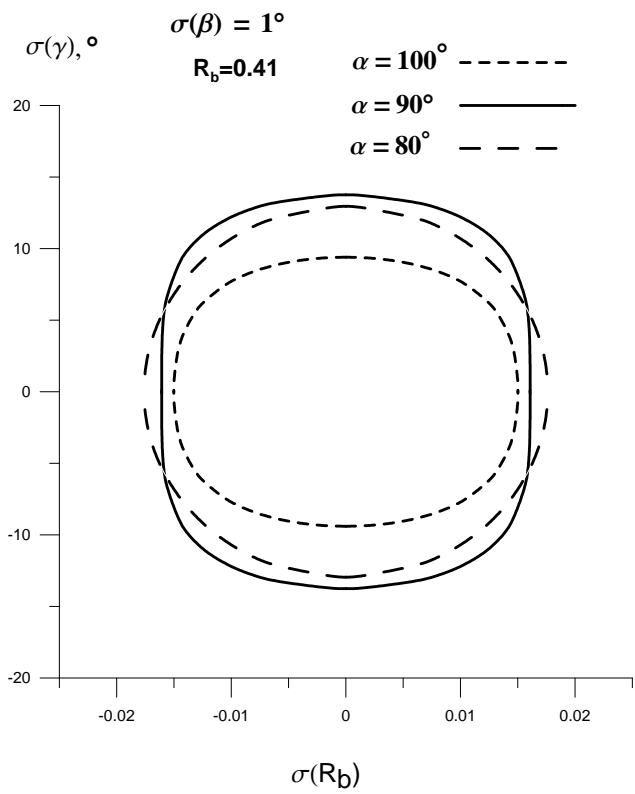} \caption{The same as Fig.1 for
different values of the angle $\alpha$.}
\end{figure}

\begin{figure}[!t]
\epsfxsize=10cm \epsfbox{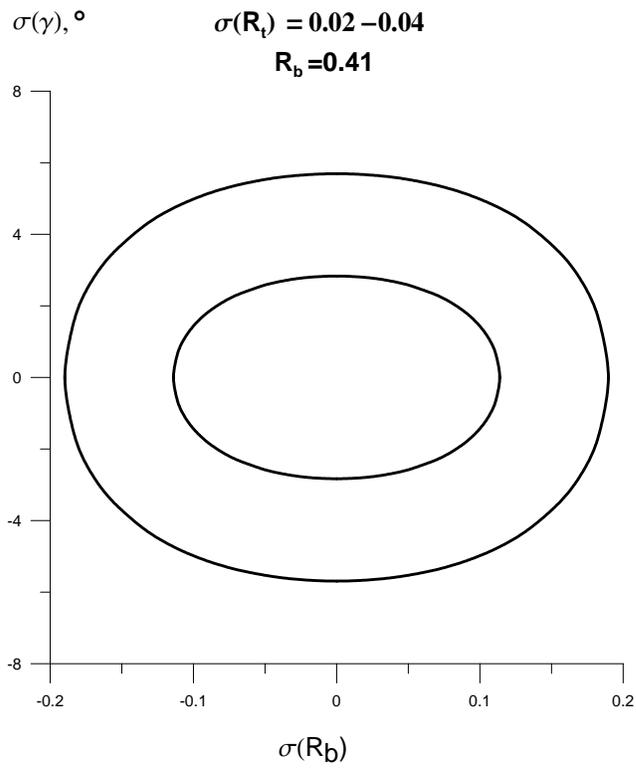} \caption{Error propagation for
$R_t$ from (\ref{rb}). The curves correspond to $\sigma (R_t) =
0.02$ and $0.04$. The choice for other parameters is $R_b = 0.41$,
$\gamma = 1$ rad. }
\end{figure}

\begin{figure}[!t]
\epsfxsize=10cm \epsfbox{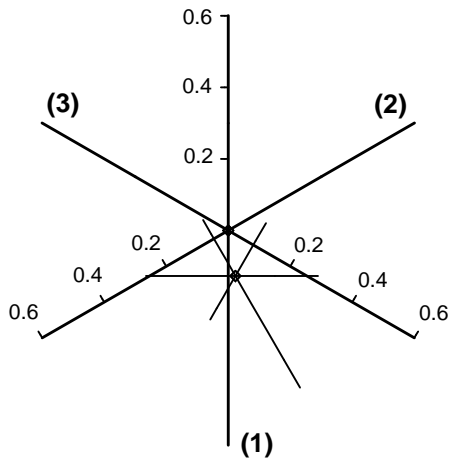} \caption{The results for $\log
\zeta_{s,K^*}^{(i)}$, the case [44]-B plotted as a point in ternary
coordinates. The SM solution is the point at the origin. The
algebraic distance from the $i$-th axis is given by $\log
\zeta_{s,K^*}^{(i)}$, positive for two axes forming an angle a point
belongs to and negative for the remaining distant axis. With this
rule each point on the plain satisfies the constraint (\ref{basic}).
The bars correspond to $1 \sigma$ deviation in $\zeta_{s,K^*}^{(i)}$
, not in $\log \zeta_{s,K^*}^{(i)}$. The marks on axes set the scale
and serve mainly for guiding eyes.}
\end{figure}

\end{document}